\documentclass[pre,aps,preprint,showpacs,preprintnumbers,amsmath,amssymb,secnumroman,eqsecnum]{revtex4}

\usepackage{graphicx}
\usepackage{dcolumn}
\usepackage{bm}

\newcommand{\beq}{\begin{equation}}
\newcommand{\eeq}{\end{equation}}
\newcommand{\beqa}{\begin{eqnarray}}
\newcommand{\eeqa}{\end{eqnarray}}
\newcommand{\vc}[1]{\mbox{\boldmath $#1$}}

\newcommand{\vol}[1]{{\bf #1}}

\newcommand{\du}[1]{{\bf\sf #1}}

\begin{document}


\title{Stokesian swimming of a sphere at low Reynolds number}

\author{B. U. Felderhof}

 \email{ufelder@physik.rwth-aachen.de}
\affiliation{Institut f\"ur Theorie der Statistischen Physik\\ RWTH Aachen University\\
Templergraben 55\\52056 Aachen\\ Germany\\
}%

\author{R. B. Jones}

 \email{r.b.jones@qmul.ac.uk}
\affiliation{Queen Mary University of London, The School of
Physics and Astronomy, Mile End Road, London E1 4NS, UK\\}%

\date{\today}

\begin{abstract}
Explicit expressions are derived for the matrices determining the mean translational and rotational swimming velocities and the mean rate of dissipation for Stokesian swimming at low Reynolds number of a distorting sphere in a viscous incompressible fluid. As an application an efficient helical propeller-type stroke is found and its properties are calculated.
\end{abstract}

\maketitle

\section{\label{I}Introduction}
Stokesian swimming of a planar sheet by means of a transverse surface wave was first studied by Taylor \cite{1}. His analysis was soon extended to a squirming sphere by Lighthill \cite{2}. Blake corrected this work and applied it in a spherical envelope approach to ciliary propulsion \cite{3}. The calculations of mean translational swimming velocity and rate of dissipation are complicated, and for simplicity the analysis was restricted to axial strokes. Recently Pedley et al. \cite{4} have included an axisymmetric azimuthal flow and showed that this can lead to a mean rotational swimming velocity. Shapere and Wilczek \cite{5} derived a general expression for the mean translational swimming velocity of a sphere, but it is not easily applied, and analogous expressions for the mean rotational swimming velocity and the mean rate of dissipation are missing.

In earlier work \cite{6} we have derived general expressions for mean translational swimming velocity and mean rate of dissipation for small amplitude swimming of a sphere in a fluid with inertia, but applied the analysis only to flow restricted to potential modes. Later we included viscous modes \cite{7}, and studied in particular the efficiency of swimming. Shapere and Wilczek \cite{8} showed that a convenient definition of efficiency leads to the formulation of an eigenvalue problem involving two matrices, one for the mean translational swimming velocity and one for the mean rate of dissipation. We derived explicit expressions for the two matrices in a convenient representation, but for simplicity restricted the analysis to axial flows, in the manner of Lighthill \cite{2} and Blake \cite{3}.

In the following we extend the analysis to the complete set of solutions of the steady state Stokes equations and derive explicit expressions for the integrals entering the matrices for mean translational and rotational swimming velocity and mean rate of dissipation. As an application we find an efficient helical propeller-type stroke and calculate its properties.

\section{\label{II}Flow equations}

In this section we recall the basic equations of the bilinear theory of swimming. We consider a sphere of radius $a$ immersed in a viscous
incompressible fluid of shear viscosity $\eta$. At low Reynolds
number and on a slow time scale the flow velocity
$\vc{v}(\vc{r},t)$ and the pressure $p(\vc{r},t)$ satisfy the
Stokes equations
\begin{equation}
\label{2.1}\eta\nabla^2\vc{v}-\nabla p=0,\qquad\nabla\cdot\vc{v}=0.
\end{equation}
The fluid is set in motion by distortions of the
spherical surface which are periodic in time and lead to swimming
motion of the sphere as well as to a time-dependent flow field. The surface displacement
$\vc{\xi}(\vc{s},t)$ is defined as the vector distance
\begin{equation}
\label{2.2}\vc{\xi}=\vc{s}'-\vc{s}
\end{equation}
of a point $\vc{s}'$ on the displaced surface $S(t)$ from the
point $\vc{s}$ on the sphere with surface $S_0$. The fluid
velocity $\vc{v}(\vc{r},t)$ is required to satisfy
\begin{equation}
\label{2.3}\vc{v}(\vc{s}+\vc{\xi}(\vc{s},t))=\frac{\partial\vc{\xi}(\vc{s},t)}{\partial t},
\end{equation}
corresponding to a no-slip boundary condition. The instantaneous translational swimming velocity $\vc{U}(t)$,
the rotational swimming velocity $\vc{\Omega}(t)$, and the flow pattern $(\vc{v},p)$ follow from the condition that no net
force or torque is exerted on the fluid. We evaluate these quantities by a perturbation expansion in powers of the
displacement $\vc{\xi}(\vc{s},t)$.

To second order in $\vc{\xi}$ the flow velocity and the swimming velocity
take the form \cite{9}
\begin{equation}
\label{2.4}\vc{v}(\vc{r},t)=\vc{v}_1(\vc{r},t)+\vc{v}_2(\vc{r},t)+...,\qquad
\vc{U}(t)=\vc{U}_2(t)+....
\end{equation}
Both $\vc{v}_1$ and $\vc{\xi}$ vary harmonically with frequency
$\omega$, and can be expressed as
 \begin{eqnarray}
\label{2.5}\vc{v}_1(\vc{r},t)&=&\vc{v}_{1c}(\vc{r})\cos\omega t+\vc{v}_{1s}(\vc{r})\sin\omega t,\nonumber\\
\vc{\xi}(\vc{s},t)&=&\vc{\xi}_{c}(\vc{s})\cos\omega
t+\vc{\xi}_{s}(\vc{s})\sin\omega t.
\end{eqnarray}
Expanding the no-slip condition Eq. (2.3) to second order we find
for the flow velocity at the surface
\begin{eqnarray}
\label{2.6}\vc{u}_{1S}(\theta,\varphi,t)&=&\vc{v}_1\big|_{r=a}=\frac{\partial\vc{\xi}(\theta,\varphi,t)}{\partial t},\nonumber\\
\vc{u}_{2S}(\theta,\varphi,t)&=&\vc{v}_2\big|_{r=a}=-\vc{\xi}\cdot\nabla\vc{v}_1\big|_{r=a},
\end{eqnarray}
in spherical coordinates $(r,\theta,\varphi)$. In complex notation with $\vc{v}_1=\vc{v}_\omega\exp(-i\omega t)$ the mean second order surface velocity is given by
\begin{equation}
\label{2.7}\overline{\vc{u}}_{2S}(\vc{s})=-\frac{1}{2}\mathrm{Re}(\vc{\xi}^*_\omega\cdot\nabla)\vc{v}_\omega\big|_{r=a},
\end{equation}
where the overhead bar indicates a time-average over a period $T=2\pi/\omega$.

We consider periodic displacements such that the body swims in the $z$ direction. The time-averaged translational swimming velocity is given by \cite{6}
\begin{equation}
\label{2.8}\overline{\vc{U}}_2=\overline{U_2}\;\vc{e}_z,\qquad\overline{U_2}=-\frac{1}{4\pi}\int\overline{\vc{u}}_{2S}\cdot\vc{e}_z\;d\Omega,
\end{equation}
where the integral is over spherical angles $(\theta,\varphi)$.
Similarly the time-averaged rotational swimming velocity is given by  \cite{6}
\begin{equation}
\label{2.9}\overline{\vc{\Omega}}_2=\overline{\Omega_2}\;\vc{e}_z,\qquad\overline{\Omega_2}=-\frac{3}{8\pi a}\int(\vc{e}_r\times\overline{\vc{u}}_{2S})\cdot\vc{e}_z\;d\Omega.
\end{equation}

To second order the rate of dissipation $\mathcal{D}_2(t)$ is
determined entirely by the first order solution. It may be
expressed as a surface integral \cite{6}
\begin{equation}
\label{2.10}\mathcal{D}_2=-\int_{r=a}\vc{v}_{1}\cdot\vc{\sigma}_{1}\cdot\vc{e}_r\;dS,
\end{equation}
where $\vc{\sigma}_1$ is the first order stress tensor, given by
\begin{equation}
\label{2.11}\vc{\sigma}_{1}=\eta(\nabla\vc{v}_1+\widetilde{\nabla\vc{v}_1})-p_1\vc{I}.
\end{equation}
The rate of dissipation is positive and oscillates in time about a
mean value. The mean rate of dissipation equals the power
necessary to generate the motion.

\section{\label{III}Matrices}

In our calculations it is convenient to expand the first order flow field and the pressure in terms of a basis set of complex solutions. The general solution of Eq. (2.1) which tends to zero at infinity and varies harmonically in time can be expressed as the complex flow velocity and pressure
\begin{eqnarray}
\label{3.1}\vc{v}^c_1(\vc{r},t)&=&-\omega a\sum^\infty_{l=1}\sum^l_{m=-l}\big[\kappa_{lm}\vc{v}_{lm}(\vc{r})+\nu_{lm}\vc{w}_{lm}(\vc{r})+\mu_{lm}\vc{u}_{lm}(\vc{r})\big]e^{-i\omega t},\nonumber\\
p^c_1(\vc{r},t)&=&-\omega a\sum^\infty_{l=1}\sum^l_{m=-l}\kappa_{lm}p_{lm}(\vc{r})e^{-i\omega t},
\end{eqnarray}
with complex coefficients $(\kappa_{lm},\nu_{lm},\mu_{lm})$ and basic solutions \cite{7},\cite{10},\cite{11}
\begin{eqnarray}
\label{3.2}
\vc{v}_{lm}(\vc{r})&=&\bigg(\frac{2l+2}{l(2l+1)}\hat{\vc{A}}_{lm}-\frac{2l-1}{2l+1}\hat{\vc{B}}_{lm}\bigg)\bigg(\frac{a}{r}\bigg)^l,\nonumber\\ p_{lm}(\vc{r})&=&2\eta(2l-1)(-1)^mP^m_l(\cos\theta)e^{im\varphi}\frac{a^l}{r^{l+1}},\nonumber\\
\vc{w}_{lm}(\vc{r})&=&-i\;\hat{\vc{C}}_{lm}\bigg(\frac{a}{r}\bigg)^{l+1},\nonumber\\
\vc{u}_{lm}(\vc{r})&=&-\hat{\vc{B}}_{lm}\bigg(\frac{a}{r}\bigg)^{l+2},
\end{eqnarray}
with vector spherical harmonics $\hat{\vc{A}}_{lm},\hat{\vc{B}}_{lm},\hat{\vc{C}}_{lm}$ in the notation of Ref. 11 (with $2^{l+1}$ in the normalization coefficient replaced by $2l+1$), and with associated Legendre functions $P^m_l$ in the notation of Edmonds \cite{12}. The functions $(\vc{v}_{lm}(\vc{r}), p_{lm}(\vc{r}))$ satisfy the Stokes equations (2.1), and the functions $\vc{u}_{lm}(\vc{r})$ and $\vc{w}_{lm}(\vc{r})$ satisfy these equations with vanishing pressure disturbance. For $m=0$ the solutions are axisymmetric and the functions  $\vc{u}_{lm}(\vc{r}),\vc{v}_{lm}(\vc{r}),p_{lm}(\vc{r})$ are then identical with the functions $\vc{u}_{l}(\vc{r}),\vc{v}_{l}(\vc{r}),p_{l}(\vc{r})$ introduced in Ref. 7. The solutions contain a factor $\exp[i(m\varphi-\omega t)]$, representing a running wave in the azimuthal direction for $m\neq0$.

We consider a superposition of solutions of the form Eq. (3.1) with a single value of $m$. The corresponding surface displacement $\vc{\xi}(\vc{s},t)$ takes the form
\begin{equation}
\label{3.3}\vc{\xi}(\vc{s},t)=-i a\sum^\infty_{l=m}\big[\kappa_{lm}\vc{v}_{lm}(\vc{s})+\nu_{lm}\vc{w}_{lm}(\vc{s})+\mu_{lm}\vc{u}_{lm}(\vc{s})\big]e^{-i\omega t}.
\end{equation}
Correspondingly we introduce the moment vector
\begin{equation}
\label{3.4}\vc{\psi}=(\kappa_{lm},\nu_{lm},\mu_{lm},\kappa_{l+1,m},...).
\end{equation}
Then the mean second order swimming velocity is in the $z$ direction with value $\overline{U_2}$ given by
\begin{equation}
\label{3.5}\overline{U_2}=\frac{1}{2}\;\omega
a(\vc{\psi}|\du{B}|\vc{\psi}),
\end{equation}
with a dimensionless hermitian matrix $\du{B}$. The mean second order rotational swimming velocity is in the $z$ direction with value $\overline{\Omega_2}$ given by
\begin{equation}
\label{3.6}\overline{\Omega_2}=\frac{3}{4}\;\omega
(\vc{\psi}|\du{C}|\vc{\psi}),
\end{equation}
with a dimensionless hermitian matrix $\du{C}$. The time-averaged rate of dissipation can be expressed as
\begin{equation}
\label{3.7}\overline{\mathcal{D}_2}=8\pi\eta\omega^2a^3(\vc{\psi}|\du{A}|\vc{\psi}),
\end{equation}
with a dimensionless hermitian matrix $\du{A}$. The matrix elements of the three matrices can be evaluated from Eqs. (2.6)-(2.11).

For $m=0,1$ we must put the moments $\kappa_{1m},\nu_{1m}$ equal to zero, because of the requirement that the swimmer exert no net force or torque on the fluid. Correspondingly for $m=0,1$ the matrices $\du{A},\du{B},\du{C}$ can be truncated by deleting the first two rows and columns. In the following we assume that this truncation has been performed.

The matrix $\du{A}$ turns out to be diagonal in the subscripts $l,m$, as given by a factor $\delta_{ll'}\delta_{mm'}$.
The basis elements $\vc{v}_{lm},\;\vc{w}_{lm},\;\vc{u}_{lm}$ will be indicated by a discrete index $\sigma$ taking the values $(0,1,2)$. Then the matrix $\du{A}$ at position $lm$ has elements of the $3\times 3$ matrix of the form
\begin{equation}
\label{3.8}\du{A}_{lm}=\left(\begin{array}{ccc}
a_{lm00}&0&a_{lm02}\\
0&a_{lm11}&0\\
\\a_{lm20}&0&a_{lm22}
\end{array}\right).
\end{equation}
The nonvanishing $00,\;02,\;20,\;22$ elements are determined by the integrals
\begin{eqnarray}
\label{3.9}\int_{r=a}\vc{u}^*_{km}\cdot(\nabla\vc{u}_{lm})\cdot\vc{e}_r\;dS&=&-4\pi af_{km}
(k+1)(k+2)\delta_{kl},\nonumber\\
\int_{r=a}\vc{u}^*_{km}\cdot(\nabla\vc{v}_{lm}+\widetilde\nabla\vc{v}_{lm}-p_{lm}/\eta)\cdot\vc{e}_r\;dS&=&-8\pi af_{km}
\frac{(k+1)(k+2)(2k-1)}{2k+1}\;\delta_{kl},\nonumber\\
\int_{r=a}\vc{v}^*_{km}\cdot(\nabla\vc{u}_{lm})\cdot\vc{e}_r\;dS&=&-4\pi af_{km}
\frac{(k+1)(k+2)(2k-1)}{2k+1}\;\delta_{kl},\nonumber\\
\int_{r=a}\vc{v}^*_{km}\cdot(\nabla\vc{v}_{lm}+\widetilde\nabla\vc{v}_{lm}-p_{lm}/\eta)\cdot\vc{e}_r\;dS&=&-8\pi af_{km}
\frac{(k+1)(2k^3+k^2-2k+2)}{k(2k+1)}\;\delta_{kl},\nonumber\\
\end{eqnarray}
with the factor
\begin{equation}
\label{3.10}f_{lm}=\frac{(l+m)!}{(l-m)!}.
\end{equation}
For $m=0$ the expressions reduce to those given in Eq. (7.14) of Ref. 7.
The $11$ integral is given by
\begin{equation}
\label{3.11}\int_{r=a}\vc{w}^*_{km}\cdot(\nabla\vc{w}_{lm}+\widetilde{\nabla\vc{w}_{lm}})\cdot\vc{e}_r\;dS=-4\pi a f_{km}\frac{k(k+1)(k+2)}{2k+1}\;\delta_{kl}.
\end{equation}

The $\du{B}$ and $\du{C}$-matrices have $6\times 6$ matrices along the diagonal and are of the form
\begin{equation}
\label{3.12}\du{B}_{lm}=\left(\begin{array}{cc}
B^{(1)}_{lm}&B^{(2)}_{lm}\\
B^{(3)}_{lm}&B^{(4)}_{lm}
\end{array}\right),
\end{equation}
with $3\times 3$ matrices $B^{(j)}_{lm}$ with the relation $B^{(4)}_{lm}=B^{(1)}_{l+1,m}$. The $\du{B}$ and $\du{C}$-matrices have a checkerboard form with zeros on alternate positions. The integrals determining the $\du{B}$-matrix are similar to those given in Eq. (7.7) of Ref. 7. Explicitly we find
\begin{eqnarray}
\label{3.13}\int_{r=a}\vc{v}^*_{km}\cdot(\nabla\vc{v}_{lm})\cdot\vc{e}_z\;dS=&-&4\pi a\bigg[g_{km}
\frac{2(k+2)(2k-1)}{(2k+1)(2k+3)}\;\delta_{k,l-1}+
g_{lm}\frac{(l+2)(2l-1)^2}{(2l+1)(2l+3)}\;\delta_{k,l+1}\bigg],\nonumber\\
\int_{r=a}\vc{v}^*_{km}\cdot(\nabla\vc{w}_{lm})\cdot\vc{e}_z\;dS=&&4\pi amf_{km}\frac{(k+1)(2k-1)}{2k+1}\;\delta_{kl},\nonumber\\
\int_{r=a}\vc{v}^*_{km}\cdot(\nabla\vc{u}_{lm})\cdot\vc{e}_z\;dS=&-&4\pi ag_{lm}\frac{(l+2)(2l+1)}{2l+3}\;\delta_{k,l+1},\nonumber\\
\int_{r=a}\vc{u}^*_{km}\cdot(\nabla\vc{v}_{lm})\cdot\vc{e}_z\;dS=&-&4\pi a\bigg[
g_{km}\frac{2(k+2)}{2k+3}\;\delta_{k,l-1}+
g_{lm}\frac{(l+2)(2l-1)}{2l+3}\;\delta_{k,l+1}\bigg],\nonumber\\
\int_{r=a}\vc{u}^*_{km}\cdot(\nabla\vc{w}_{lm})\cdot\vc{e}_z\;dS&=&4\pi a
mf_{km}(k+1)\delta_{kl},\nonumber\\
\int_{r=a}\vc{u}^*_{km}\cdot(\nabla\vc{u}_{lm})\cdot\vc{e}_z\;dS=&-&4\pi a
g_{lm}(l+2)\delta_{k,l+1},
\end{eqnarray}
with factor
\begin{equation}
\label{3.14}g_{lm}=\frac{(l+m+1)!}{(l-m)!}.
\end{equation}
The integrals determining the $\du{C}$-matrix are given by
\begin{eqnarray}
\label{3.15}\int_{r=a}[\vc{e}_r\times(\vc{v}^*_{km}\cdot\nabla\vc{v}_{lm})]\cdot\vc{e}_z\;dS&=&4\pi iamf_{km}
\frac{(k-2)(2k^3+k^2-2k+2)}{k^2(2k+1)}\;\delta_{kl},\nonumber\\
\int_{r=a}[\vc{e}_r\times(\vc{v}^*_{km}\cdot\nabla\vc{w}_{lm})]\cdot\vc{e}_z\;dS&=&4\pi ia\bigg[ g_{km}
\frac{(k+2)^2(2k-1)}{(2k+1)(2k+3)}\;\delta_{k,l-1}\nonumber\\&-&g_{lm}\frac{l(l+2)(2l^2+l+1)}{(l+1)(2l+1)(2l+3)}\delta_{k,l+1}\bigg],\nonumber\\
\int_{r=a}[\vc{e}_r\times(\vc{v}^*_{km}\cdot\nabla\vc{u}_{lm})]\cdot\vc{e}_z\;dS&=&4\pi iamf_{km}\frac{(k+2)(2k-1)}{2k+1}\;\delta_{kl},\nonumber\\
\int_{r=a}[\vc{e}_r\times(\vc{w}^*_{km}\cdot\nabla\vc{v}_{lm})]\cdot\vc{e}_z\;dS&=&-4\pi ia\bigg[
g_{km}\frac{2k(k+2)}{(k+1)(2k+1)(2k+3)}\;\delta_{k,l-1}\nonumber\\&-&g_{lm}
\frac{(l+2)(2l-1)}{(2l+1)(2l+3)}\;\delta_{k,l+1}\bigg],\nonumber\\
\int_{r=a}[\vc{e}_r\times(\vc{w}^*_{km}\cdot\nabla\vc{w}_{lm})]\cdot\vc{e}_z\;dS&=&-4\pi iamf_{km}\frac{1}{2k+1}\;\delta_{kl},\nonumber\\
\int_{r=a}[\vc{e}_r\times(\vc{w}^*_{km}\cdot\nabla\vc{u}_{lm})]\cdot\vc{e}_z\;dS&=&4\pi iag_{lm}
\frac{l+2}{2l+3}\;\delta_{k,l+1},\nonumber\\
\int_{r=a}[\vc{e}_r\times(\vc{u}^*_{km}\cdot\nabla\vc{v}_{lm})]\cdot\vc{e}_z\;dS&=&4\pi iamf_{km}
\frac{(k^2-2)(2k-1)}{k(2k+1)}\;\delta_{kl},\nonumber\\
\int_{r=a}[\vc{e}_r\times(\vc{u}^*_{km}\cdot\nabla\vc{w}_{lm})]\cdot\vc{e}_z\;dS&=&4\pi ia\bigg[g_{km}
\frac{(k+2)^2}{2k+3}\;\delta_{k,l-1}-g_{lm}\frac{l(l+2)}{2l+3}\delta_{k,l+1}\bigg],\nonumber\\
\int_{r=a}[\vc{e}_r\times(\vc{u}^*_{km}\cdot\nabla\vc{u}_{lm})]\cdot\vc{e}_z\;dS&=&4\pi
iamf_{km}(k+2)\delta_{kl}.
\end{eqnarray}
The matrices $\du{A}$ and $\du{C}$ are real and symmetric. The matrix $\du{B}$ is pure imaginary and antisymmetric. In our earlier work \cite{7} we have given explicit expressions for the upper left hand corners of the matrices $\du{A}$ and $\du{B}$ for $m=0$, before deletion of the first two rows and columns. The explicit form of the truncated low order matrices for $m=1$ is given in an example of swimming by helical wave \cite{13}.

\section{\label{IV}Eigenvalue problem}

It is of interest to optimize the mean translational swimming velocity $\overline{U}_2$ for given mean rate of dissipation $\overline{\mathcal{D}}_2$. The optimization leads to a generalized eigenvalue problem of the form
\begin{equation}
\label{4.1}\du{B}|\vc{\psi}_\lambda)=\lambda\du{A}|\vc{\psi}_\lambda).
\end{equation}
The maximum eigenvalue $\lambda_{max}$ determines the optimum translational swimming velocity. The corresponding mean rotational swimming velocity $\overline{\Omega}_2$ can be found from the eigenvector $|\psi_\lambda)$ by use of Eq. (3.6). In earlier work \cite{7} we have shown that the maximum eigenvalue for $m=0$ is given by $\lambda_{max}=2\sqrt{2}$, but this obtains in the limit where moments of arbitrarily high number $l$ are involved. The absolute value of the elements of the corresponding eigenvector tends to a constant for large $l$. Presumably the corresponding fine detail of the surface displacement is not physically relevant. In the following we consider finite-dimensional moment vectors, with $l$ at most equal to a bounded value $L$. The corresponding matrices are $(3L-3m+3)\times(3L-3m+3)$-dimensional for $m\geq2$. For $m=0,1$ the matrices are $(3L-2)\times(3L-2)$-dimensional. We denote the maximum eigenvalue with maximum $l$ equal to $L$ as $\lambda_{Lm,max}$.

In Fig. 1 we plot $\lambda_{Lm,max}$ for $L=10$ as a function of $m$. The maximum eigenvalue is maximum at $m=0$, corresponding to axial symmetry. We find $\lambda_{10,0,max}=2.348$, not much less than $2\sqrt{2}\approx 2.828$. For $m=1$ the value is $\lambda_{10,0,max}=2.344$, and for $m=10$ it is $\lambda_{10,10,max}=1.708$.

In the eigenvector at $m=0$ the elements $\nu_{l0}$ vanish, so that in optimum swimming the rotational modes $\vc{w}_{l0}$ with swirl are absent. The low order truncated matrix $\du{C}_{20}$ corresponding to $L=2,m=0$ reads explicitly
\begin{equation}
\label{4.2}\du{C}_{20}=\frac{12}{5}\left(\begin{array}{cccc}
0&0&1&0
\\0&0&0&0
\\1&0&0&0
\\0&0&0&0
\end{array}\right).
\end{equation}
This shows that a moment vector with nonvanishing $\kappa_{10}$ and $\nu_{20}$ elements leads to a nonvanishing $\overline{\Omega}_2$. The elements correspond to the integrals for the pairs $\vc{v}_{10},\vc{w}_{20}$ and $\vc{w}_{20},\vc{v}_{10}$ in Eq. (3.15). Clearly there are similar couplings for $l>2$. These couplings explain the nonvanishing $\overline{\Omega}_2$ in the study of Pedley et al. \cite{4} for the axisymmetric case $m=0$.

In earlier work \cite{7} we have shown the absolute values of the moments $\kappa_{l0},\mu_{l0}$ of the optimal eigenvector for $L=7,m=0$. We also showed the end of the displacement vector $\vc{\xi}(\theta,0,t)$ at various angles $\theta$ in the meridional plane $\varphi=0$. In Fig. 2 we show the real and imaginary parts of the radial displacement $\xi_r(\theta,0,0)$, and in Fig. 3 we show the real and imaginary parts of the component $\xi_\theta(\theta,0,0)$ for the optimal eigenvector for $L=10,m=0$. The eigenvector has been normalized to unity and its first component is chosen to be positive. The tangential component of $\vc{\xi}(\theta,0,0)$ is about twice as large as the radial component.

For $m\neq 0$ the rotational swimming velocity does not vanish for the eigenvector $\vc{\xi}_{Lm}$ corresponding to $\lambda_{Lm,max}$. In Fig. 4 we plot the ratio
\begin{equation}
\label{4.3}\rho_{Lm}=\frac{(\vc{\xi}_{Lm}|\du{C}_{Lm}|\vc{\xi}_{Lm})}{(\vc{\xi}_{Lm}|\du{A}_{Lm}|\vc{\xi}_{Lm})}
\end{equation}
as a function of $m$ for $L=10$.  This shows the mean rate of rotation for optimal strokes with maximum $l$ number equal to $L=10$ and different $m$, normalized to equal power.
In particular, at $m=1$ the ratio is $\rho_{10,1}=0.077$, and at $m=10$ the ratio is $\rho_{10,10}=0.894$. At $m=10$ the eigenvector is $\vc{\xi}_{10,10}=(0,1,-0.488i)$ with nonvanishing moments $(\nu_{10,10},\mu_{10,10})$.

The last example corresponds to a propeller-type first order flow. We consider more generally the case $m=L$. The matrices $\du{A}_{LL},\du{B}_{LL},\du{C}_{LL}$ are 3-dimensional and the eigenvalue equation can be solved in analytic form. The matrix $\du{A}_{LL}$ is given by
\begin{equation}
\label{4.4}\du{A}_{LL}=\frac{(L+1)(2L-1)!}{2L+1}\left(\begin{array}{ccc}
2L^3+L^2-2L+2&0&L(L+2)(2L-1)
\\0&\frac{1}{2}L^2(L+2)&0
\\L(L+2)(2L-1)&0&L(L+2)(2L+1)
\end{array}\right).
\end{equation}
The matrix $\du{B}_{LL}$ is given by
\begin{equation}
\label{4.5}\du{B}_{LL}=i\frac{L(L+1)(2L)!}{2(2L+1)}\left(\begin{array}{ccc}
0&-2L+1&0
\\2L-1&0&2L+1
\\0&-2L-1&0
\end{array}\right).
\end{equation}
The matrix $\du{C}_{LL}$ is given by
\begin{equation}
\label{4.6}\du{C}_{LL}=\frac{(2L)!}{L(2L+1)}\left(\begin{array}{ccc}
(L-2)(2L^3+L^2-2L+2)&0&L(2L^3+L^2-3L+1)
\\0&-L^2&0
\\L(2L^3+L^2-3L+1)&0&L^2(L+2)(2L+1)
\end{array}\right).
\end{equation}
We find for the maximum eigenvalue of the problem Eq. (4.1)
\begin{equation}
\label{4.7}\lambda_{LL}=\frac{\sqrt{2L(2L+1)}}{L+2}.
\end{equation}
This tends to $2$ as $L\rightarrow\infty$, indicating an efficient swimmer. The corresponding eigenvector is
\begin{equation}
\label{4.8}\vc{\xi}_{LL}=(0,1,\frac{-iL}{\sqrt{2L(2L+1)}}),
\end{equation}
and the ratio $\rho_{LL}$ is
\begin{equation}
\label{4.9}\rho_{LL}=\frac{L^2+2L-2}{L^2+3L+2}.
\end{equation}
We denote the reduced power of the optimal swimmer with moments $\vc{\xi}_{LL}$ as $P_{LL}$,
\begin{equation}
\label{4.10}P_{LL}=(\vc{\xi}_{LL}|\du{A}_{LL}|\vc{\xi}_{LL})=\frac{L(L+1)(L+2)}{4L+2}(2L)!.
\end{equation}
The time the swimmer needs to move over a distance equal to one diameter is \cite{14}
\begin{equation}
\label{4.11}t_{LL}=\frac{2a}{\overline{U}_2}=\frac{4}{\omega\lambda_{LL}P_{LL}}.
\end{equation}
During this time the swimmer rotates over  the angle
\begin{equation}
\label{4.12}\overline{\Omega}_2t_{LL}=\frac{3\rho_{LL}}{\lambda_{LL}}.
\end{equation}
This is independent of the power, tends to $3/2$ as $L\rightarrow\infty$, and has a maximum $(61/8)\sqrt{3/70}\approx 1.5785$ at $L=7$.

In Fig. 5 we show the imaginary part of the radial displacement $\xi_r(\theta,0,0)$ for the eigenvector $\vc{\xi}_{10,10}=(0,1,-0.488i)$ for $L=10,m=10$. The real part vanishes.
In Fig. 6 we show the real and imaginary parts of the component $\xi_\theta(\theta,0,0)$. In Fig. 7 we show the real and imaginary parts of the component $\xi_\varphi(\theta,0,0)$. It is striking that these components have a simple dependence on the polar angle $\theta$, concentrating the displacement along the equator. In the azimuthal direction there is a running wave given by the factor $\exp[i(m\varphi-\omega t)]$ with $m=10$.

\section{\label{II}Discussion}

The calculation provides the complete set of matrix elements required in the discussion of translational and rotational swimming of a distorting sphere immersed in a viscous incompressible fluid, with no-slip boundary condition on the surface of the body. The calculation is in the spirit of Lighthill \cite{2} and Blake {3}, but allows general surface distortion. The last example of Sec. IV with a propeller type azimuthal surface wave demonstrates the power of the method.

One can imagine a wide range of possible surface waves. The optimization of translational swimming velocity for given power leads to an eigenvalue problem allowing selection of the best stroke. Structural considerations can lead to constraints imposed on the nature of the stroke. For example, one can require the surface displacement to be tangential \cite{13} or radial \cite{14}.

The spherical geometry has the advantage that the complete set of solutions of the Stokes equations is known in terms of elementary functions. The calculation can serve as a guide to the study of more complicated geometry, for example with the goal of finding the optimal swimming modes of a spheroid.

\newpage

\section*{Figure captions}

\subsection*{Fig. 1}
Plot of the maximum eigenvalue $\lambda_{max}(10,m)$ as a function of $m$. This characterizes the mean translational swimming velocity
of the optimal swimmer for $L=10$ at each value of $m$ for given power.

\subsection*{Fig. 2}
Plot of the real (solid curve) and imaginary (dashed curve) parts of the radial displacement $\xi_r$ in the meridional plane $\varphi=0$ at time $t=0$ as a function of polar angle $\theta$ for the optimal swimmer with $L=10,\;m=0$.

\subsection*{Fig. 3}
Plot of the real (solid curve) and imaginary (dashed curve) parts of the tangential displacement $\xi_\theta$ in the meridional plane $\varphi=0$ at time $t=0$ as a function of polar angle $\theta$ for the optimal swimmer with $L=10,\;m=0$.

\subsection*{Fig. 4}
Plot of the ratio $\rho_{Lm}$, defined in Eq. (4.3), for $L=10$ as a function of $m$. This characterizes the rate
of steady rotation of the optimal swimmer for $L=10$ at each value of $m$ for given power.

\subsection*{Fig. 5}
Plot of the radial displacement $\xi_r$ in the meridional plane $\varphi=0$ at time $t=0$ as a function of polar angle $\theta$ for the optimal swimmer with $L=10,\;m=10$.

\subsection*{Fig. 6}
Plot of the real (solid curve) and imaginary (dashed curve) parts of the tangential displacement $\xi_\theta$ in the meridional plane $\varphi=0$ at time $t=0$ as a function of polar angle $\theta$ for the optimal swimmer with $L=10,\;m=10$.

\subsection*{Fig. 7}
Plot of the real (solid curve) and imaginary (dashed curve) parts of the azimuthal displacement $\xi_\varphi$ in the meridional plane $\varphi=0$ at time $t=0$ as a function of polar angle $\theta$ for the optimal swimmer with $L=10,\;m=10$.

\newpage
\clearpage
\newpage
\setlength{\unitlength}{1cm}
\begin{figure}
 \includegraphics{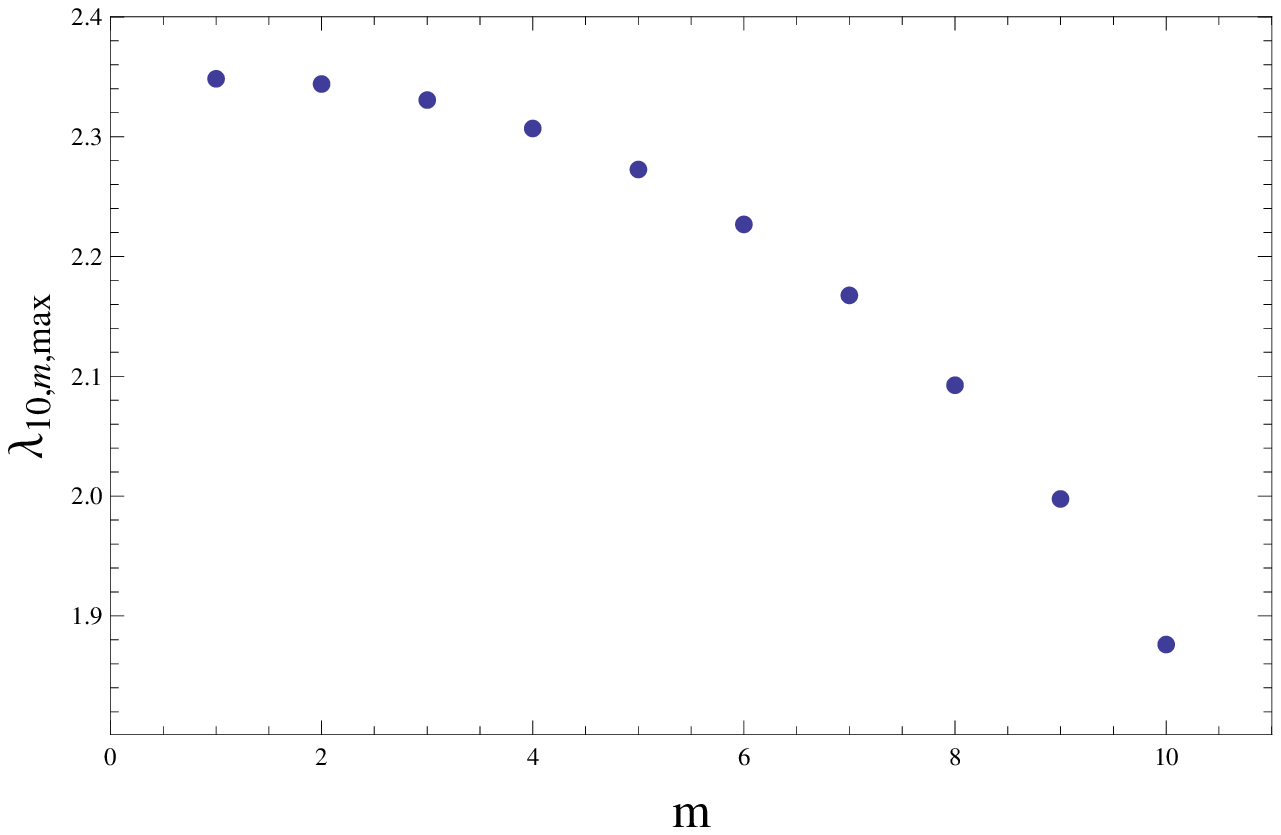}
   \put(-9.1,3.1){}
\put(-1.2,-.2){}
  \caption{}
\end{figure}
\newpage
\newpage
\clearpage
\newpage
\setlength{\unitlength}{1cm}
\begin{figure}
 \includegraphics{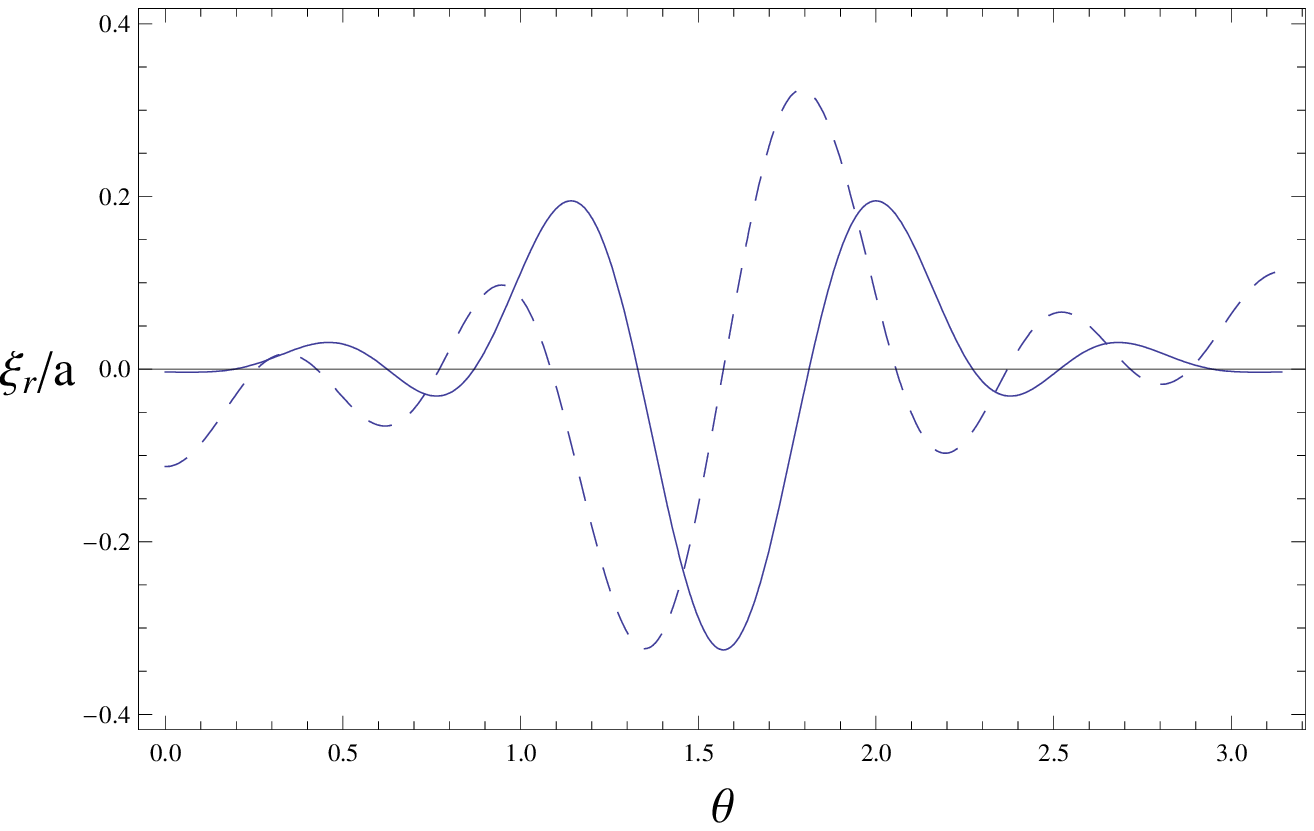}
   \put(-9.1,3.1){}
\put(-1.2,-.2){}
  \caption{}
\end{figure}
\newpage
\newpage
\clearpage
\newpage
\setlength{\unitlength}{1cm}
\begin{figure}
 \includegraphics{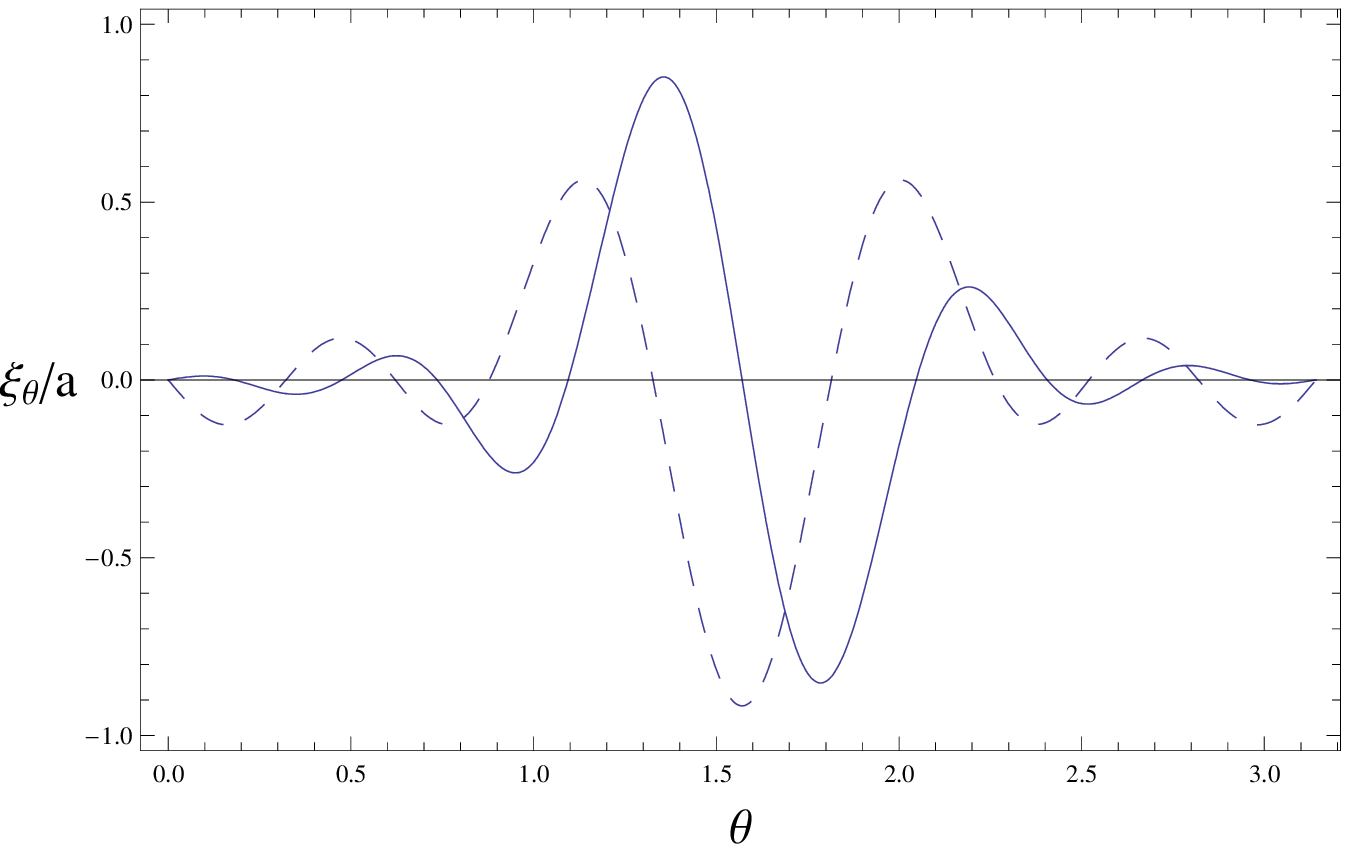}
   \put(-9.1,3.1){}
\put(-1.2,-.2){}
  \caption{}
\end{figure}
\newpage
\newpage
\clearpage
\newpage
\setlength{\unitlength}{1cm}
\begin{figure}
 \includegraphics{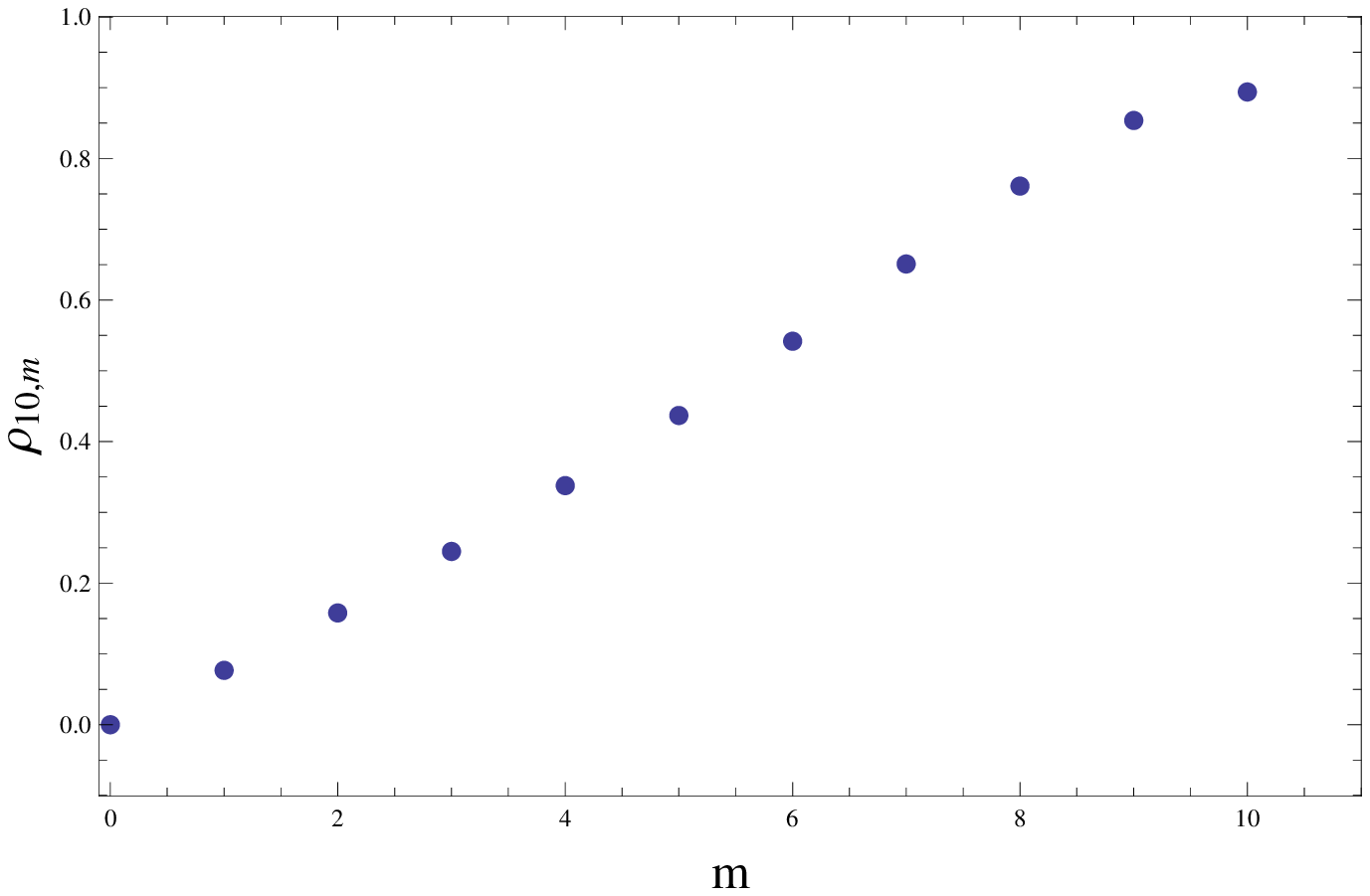}
   \put(-9.1,3.1){}
\put(-1.2,-.2){}
  \caption{}
\end{figure}
\newpage
\newpage
\clearpage
\newpage
\setlength{\unitlength}{1cm}
\begin{figure}
 \includegraphics{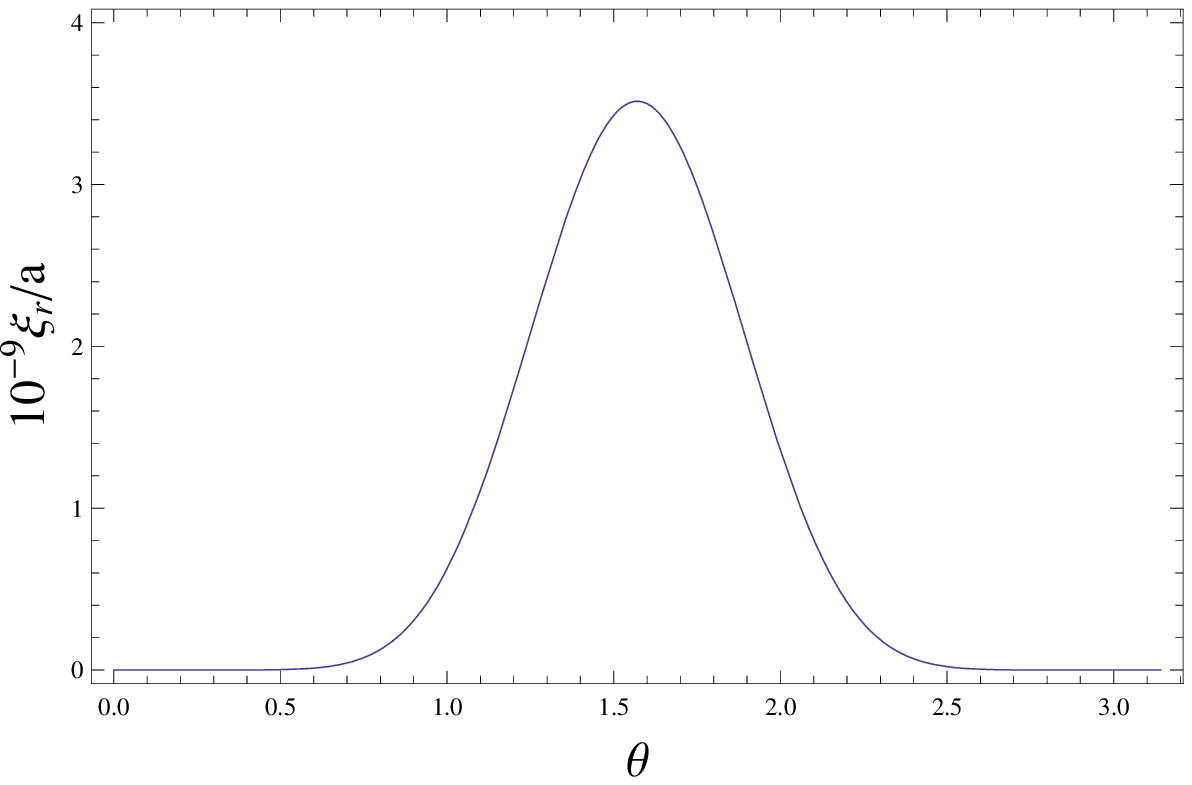}
   \put(-9.1,3.1){}
\put(-1.2,-.2){}
  \caption{}
\end{figure}
\newpage
\newpage
\clearpage
\newpage
\setlength{\unitlength}{1cm}
\begin{figure}
 \includegraphics{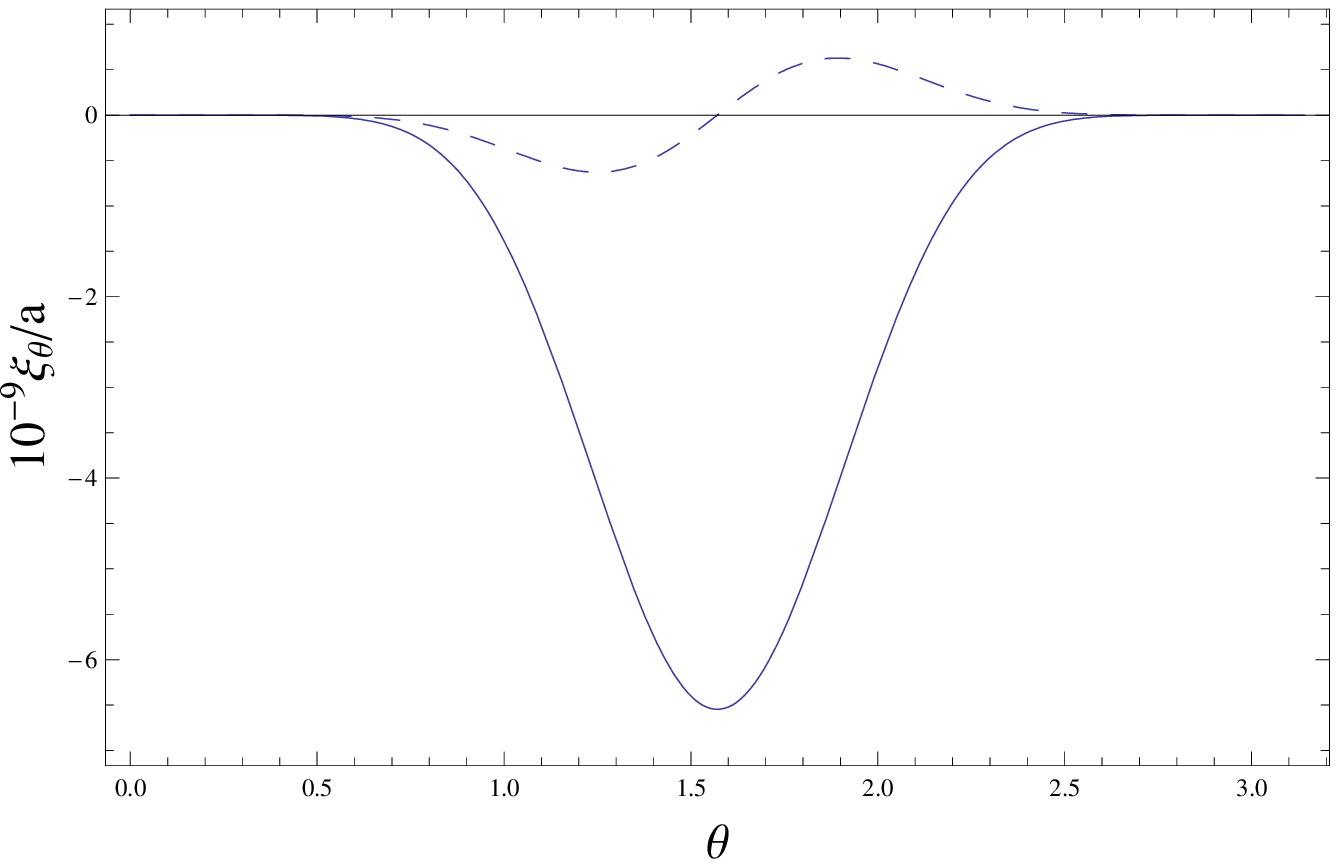}
   \put(-9.1,3.1){}
\put(-1.2,-.2){}
  \caption{}
\end{figure}
\newpage
\newpage
\clearpage
\newpage
\setlength{\unitlength}{1cm}
\begin{figure}
 \includegraphics{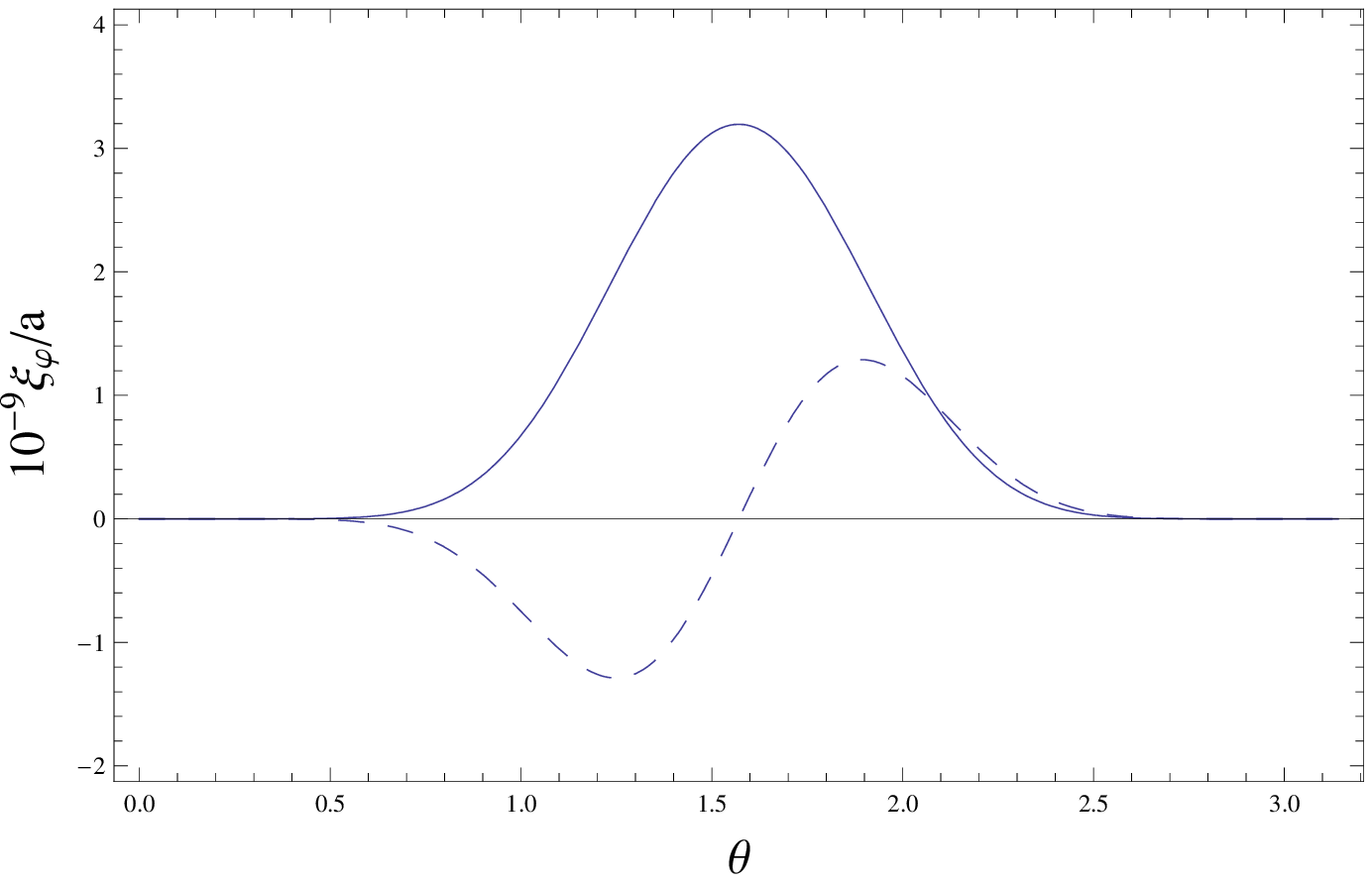}
   \put(-9.1,3.1){}
\put(-1.2,-.2){}
  \caption{}
\end{figure}
\newpage


\newpage


\begin{thebibliography}{99}

\bibitem{1}
G. I. Taylor, Analysis of the swimming of microscopic organisms, Proc. R. Soc. London A \vol{209}, 447 (1951).

\bibitem{2}
M. J. Lighthill, On the squirming motion of nearly spherical deformable bodies through liquids at very small Reynolds numbers, Comm. Pure Appl. Math. \vol{5}, 109 (1952).

\bibitem{3}
J. R. Blake, A spherical envelope approach to ciliary propulsion, J. Fluid Mech. \vol{49}, 209 (1971).

\bibitem{4}
T. J. Pedley, D. R. Brumley, and R. E. Goldstein, Squirmers with swirl - a model for Volvox swimming, arXiv:1512.02931[cond-mat.soft].

\bibitem{5}
A. Shapere and F. Wilczek, Geometry of self-propulsion at low Reynolds number, J. Fluid Mech. \vol{198}, 557 (1989).

\bibitem{6}
B. U. Felderhof and R. B. Jones, Small-amplitude swimming of a sphere, Physica A \vol{202}, 119 (1994).

\bibitem{7}
B. U. Felderhof and R. B. Jones, Optimal translational swimming of a sphere at low Reynolds number, Phys. Rev. E \vol{90}, 023008 (2014).

\bibitem{8}
A. Shapere and F. Wilczek, Efficiencies of self-propulsion at low Reynolds number, J. Fluid Mech. \vol{198}, 587 (1989).

\bibitem{9}
B. U. Felderhof and R. B. Jones, Inertial effects in small-amplitude swimming of a finite body, Physica A \vol{202}, 94 (1994).

\bibitem{10}
R. Schmitz and B. U. Felderhof, Creeping flow about a spherical particle, Physica A \vol{113}, 90 (1982).

\bibitem{11}
B. Cichocki, B. U. Felderhof, and R. Schmitz, Hydrodynamic interactions between two spherical particles, PhysicoChem. Hyd. \vol{10}, 383 (1988).

\bibitem{12}
A. R. Edmonds, {\it Angular Momentum in Quantum Mechanics} (Princeton University Press, Princeton, NJ, 1974).

\bibitem{13}
B. U. Felderhof, Spinning swimming of Volvox by tangential helical wave, arXiv:1601.00755[physics.flu-dyn].

\bibitem{14}
B. U. Felderhof, Stokesian swimming of a sphere by radial helical surface wave, arXiv:1601.03151[physics.flu-dyn].











\end{thebibliography}
\end{document}